\documentclass{vgtc}                          

\ifpdf
  \pdfoutput=1\relax                   
  \usepackage{graphicx}                
  \DeclareGraphicsExtensions{.pdf,.png,.jpg,.jpeg} 
\else
  \usepackage{graphicx}                
  \DeclareGraphicsExtensions{.eps}     
\fi%

\graphicspath{{figures/}{pictures/}{images/}{./}} 

\usepackage{microtype}                 
\PassOptionsToPackage{warn}{textcomp}  
\usepackage{textcomp}                  
\usepackage{mathptmx}                  
\usepackage{times}                     
\usepackage{cite}                      
\usepackage{tabu}                      
\usepackage{booktabs}                  
\usepackage{rotating}
\usepackage{tikz}

\usepackage{graphicx}
\usepackage{xcolor}
 
\onlineid{0}

\vgtccategory{Research}

\vgtcinsertpkg

\title{TimePool: Visually Answer ``Which and When'' Questions On Univariate Time Series}

\author{Tinghao Feng\thanks{e-mail: fengt@appstate.edu}\\ %
        \scriptsize Appalachian State University %
\and Yueqi Hu\thanks{e-mail: hyq825@gmail.com}\\ %
     \scriptsize UNC Charlotte %
\and Jing Yang\thanks{e-mail: jyang13@uncc.edu}\\ %
     \scriptsize UNC Charlotte %
\and Tom Polk\thanks{e-mail: tpolk@compassdraw.com}\\ %
     \scriptsize Compass Draw, LLC %
\and Ye Zhao\thanks{e-mail: zhao@cs.kent.edu}\\ %
     \scriptsize Kent State University %
\and Shixia Liu\thanks{e-mail: shixia@tsinghua.edu.cn}\\ %
     \scriptsize Tshinghua University %
\and Zhaocong Yang\thanks{e-mail: zyang19@uncc.edu}\\ %
     \scriptsize UNC Charlotte}

\teaser{
  \centering
  \includegraphics[width=\linewidth]{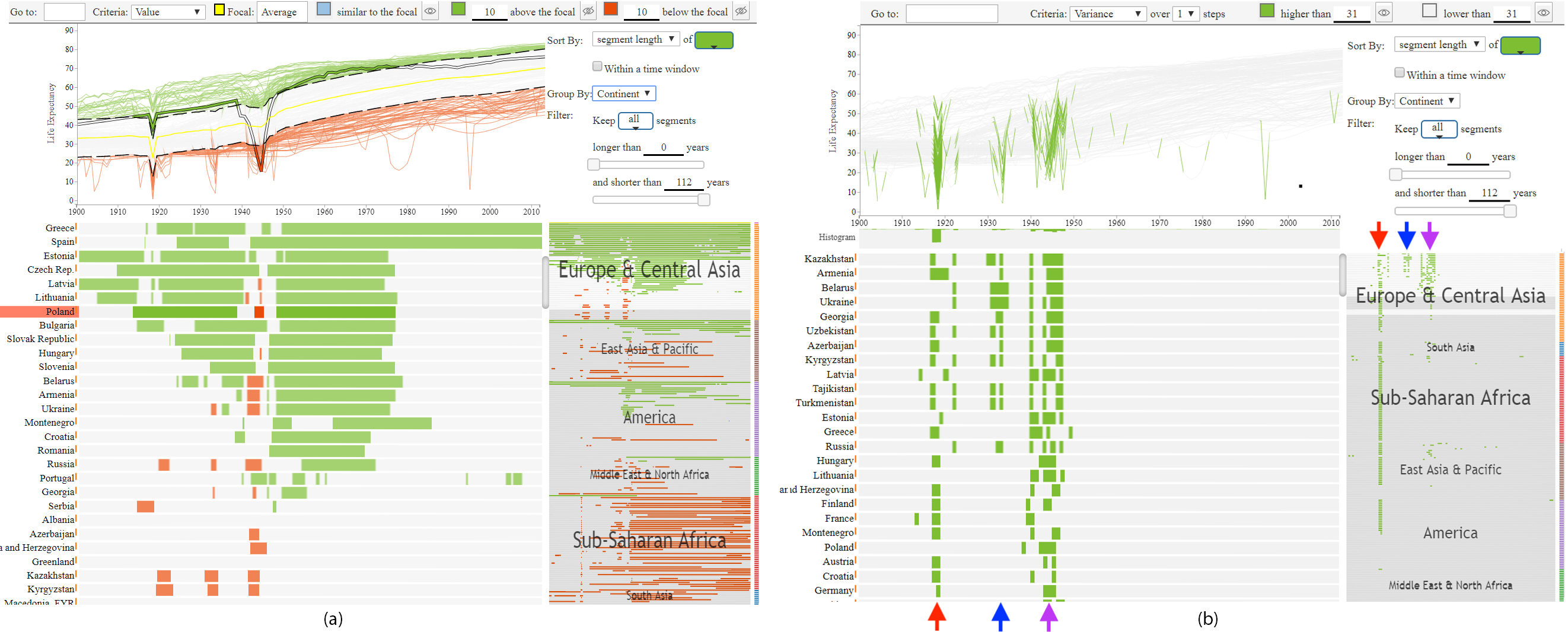}
  \caption{The World Life Expectancy dataset \cite{rosling_gapminder_2006} in TimePool. In each figure, the top left is a line chart; the bottom left is a detail view consisting of juxtaposed timelines; and the bottom right is an overview, a zoomed-out version of the detail view. Each timeline represents a country. (a) The answers to ``which countries have life expectancies 10 years longer/shorter than average? and when'' are highlighted in green/red. It can be seen from the overview that most countries in the low end are in Sub-Saharan Africa. Poland attracted attention due to its green-red-green pattern. Through the line chart, its detail can be examined. (b) The answers to ``Which countries experienced significant variances in life expectancy? and when?'' are highlighted in green. Three peaks (indicated by arrows) in the number of countries that experienced significant changes in life expectancy can be seen. The Spanish flu and World Wars occurred during those time periods.}
  \label{fig_le}
}

\abstract{When exploring time series datasets, analysts often pose ``which and when'' questions. For example, with world life expectancy data over one hundred years, they may inquire about the top 10 countries in life expectancy and the time period when they achieved this status, or which countries have had longer life expectancy than Ireland and when. This paper proposes TimePool, a new visualization prototype, to address this need for univariate time series analysis. It allows users to construct interactive ``which and when'' queries and visually explore the results for insights.} 

\CCScatlist{time series data, interaction, dynamic queries}

\begin{document}


\firstsection{Introduction}
\label{intro}
\maketitle

Nowadays, an ever-larger body of time series datasets is generated from social studies, financial activities, scientific measurements, and other applications \cite{fu2011review}. 
Among these time series, univariate time series is a typical one that consists of time-oriented data associated with a single data value.
Despite extensive research efforts, performing primitive analysis tasks visually on univariate time series remains a challenge. 
To demonstrate this, we consider the primitive analysis task of find-extremum \cite{amar2005low,andrienko2006exploratory}. When applying this task to univariate time series data, the extremum we want to find is not constant. For example, with a World Life Expectancy (WLE) dataset (200 countries, 1900-2012) \cite{rosling_gapminder_2006}, a person may ask, “Which countries were among the top 10 in life expectancy?” Since life expectancy changes over time, the answer also evolves over time, which provides rich information. For example, a person may be interested in finding countries consistently having the highest ranks, countries joining/leaving the top 10 in recent years, or countries ranked the highest during a time period of interest. 

This paper presents TimePool, a novel prototype to visually answer “which and when” questions for the following tasks:

\noindent {\bf Extreme:} Questions about extremes, namely top/bottom N data cases with respect to their original or derived values at each moment. An example question is, “Which countries were among the top 10 in life expectancy? and when?” (Q1). 

\noindent {\bf Condition:} Questions about data cases satisfying given conditions regarding their original or derived values at each moment. An example question is, “Which countries had life expectancies shorter than 50 years? and when?” (Q2). 

\noindent {\bf Comparison:} Questions about data cases satisfying given conditions when compared with other data cases using original or derived values at each moment. An example question is, “Which countries had longer life expectancies than Ireland? and when?” (Q3).

The idea of TimePool is simple: to complement a line chart with dynamic queries and views consisting of juxtaposed timelines, where users can visually and interactively explore “which and when” answers to the queries without clutter (see Fig. \ref{fig_le} and Fig. \ref{fig_compare}). For example, Fig. \ref{fig_compare}(a) presents the answers to Q1: A line chart/timeline segment is green-colored when the country was among the top 10 in life expectancy during the corresponding period. Rank is a derived value used for this query. The countries are sorted in the timeline views based on how long they stayed in the top 10. Fig. \ref{fig_compare} (c) presents the answers to Q2: A line chart/timeline segment is red-colored when the country had a life expectancy shorter than 50 years. In the timeline views, the countries are first sorted by continent and then by how long they had life expectancies shorter than 50 years between 2000 and 2012. Fig. \ref{fig_compare}(c) answers Q3 by coloring countries with longer/shorter life expectancies than Ireland in green/red in the corresponding time periods.

\section{Related Work}
\label{rela}

Line charts \cite{playfair1801commercial} display univariate time series as series of data points connected by straight lines. Comparing and following individual data cases over a wide time span is difficult \cite{javed2010graphical,ondov2018face,gogolou2018comparing}. Line chart variations, such as 3D line graphs \cite{chittaro2003data} and the braided graph \cite{javed2010graphical}, are also unintuitive and misleading with more data cases. ChronoLenses \cite{zhao2011exploratory} apply transformations (e.g. 1st derivative) on data and display the transformation results in line charts, which are usually more cluttered than the original data, for more analytical power.

Juxtaposed techniques, such as the line graph explorer \cite{kincaid2006line}, two-tone pseudo coloring \cite{saito2005two}, Spark Cloud \cite{lee2010sparkclouds}, and Horizon Graphs \cite{heer2009sizing}, display time series side by side to avoid overlapping. Interactive Horizon Graphs \cite{perin2013interactive} allow users to interactively adjust the baseline of the horizon graph to change colors and shapes for pattern discovery.  Qualizon graphs \cite{federico2014qualizon} is a variation of Horizon graphs \cite{heer2009sizing} that tie the bands with qualitative categories. Bade et al. \cite{bade2004connecting} display a time series in several levels of detail and abstraction, including both quantitative value and qualitative meaning representations. The latter is tied to \textit{a priori} or associated knowledge. Juxtaposed techniques become less effective when the number of time series displayed increases due to the intensive cognitive efforts required to decode the colors and shapes \cite{javed2010graphical, schloss2018mapping, liu2020design}. 

A timeline uses a line or a bar to represent an event's starting point and duration along a time axis \cite{aigner2011visualization}. Juxtaposed timelines are widely used for exploring events in applications such as news analysis \cite{krstajic2011cloudlines}, communication analysis \cite{fischer2021commaid, larrea2021visualization}, and medical research \cite{monroe2013temporal,klimov2010intelligent2,fails2006visual}. TimePool uses juxtaposed timelines to facilitate analyses of univariate time series, which is a quite different usage scenario from existing works. Sorting is often provided in matrix or line-based visualizations \cite{harrison1994timelines, atman2019design}. For example, matrix-based graph visualizations \cite{elmqvist2008zame} sort the matrix to reveal patterns in a graph topology. LineUp \cite{gratzl2013lineup} sorts horizontal bars representing multidimensional data to provide a ranked list. TimePool also provides sorting in its timeline views.

Dynamic queries refer to queries built dynamically with a continuous manual adjustment of one or more data values \cite{ahlberg1994visual,buono2007similarity, heer2012interactive}. This interaction has been applied to time series analysis for addressing ``which and when'' questions. Ryall et al. proposed QueryLines \cite{ryall2005querylines} to dynamically retrieve time series with a given pattern from a large amount of data. But it lacks support for ``when'' questions since determining how the pattern is distributed on the time axis is hard. Similarly, Holz et al. proposed a technique of relaxed selection \cite{holz2009relaxed}, which locates the time period that fits a given pattern under a controllable tolerance range. This approach supports the ``when'' question well but works on a single time series with no “which” question attached. Even if it answers the ``when'' question, it does not convey more details about the query since there is no additional interface to display the query results. In comparison, users can drag the threshold curve with TimePool to adjust the query condition dynamically. A separate interface helps TimePool handle a group of time series and show more details about the query results.

Exploring temporally evolving ranking of data cases in time series is challenging \cite{pereira2020rankviz}. Multiple efforts have been made to address this problem, such as Rank Chart \cite{brinton1919graphic}, VIZ-RANK \cite{perin2014table}, and Gap Chart \cite{perin2016using}. All these approaches quickly get cluttered when the number of time series displayed increases. RankExplorer \cite{shi2012rankexplorer} is scalable since it partitions a large number of time series into a manageable number of ranking categories and visually depicts the content changes between the categories over time. However, tracking the ranking evolutions of individual time series is still difficult. Instead, TimePool converts the ranking visualization problem to the question, ``Which data cases fall into a given ranking range?  and when?''. It allows users to explore the behaviors of individual data cases regarding one or more ranking ranges in a less cluttered visualization. 

\section{TimePool}


The interface of TimePool consists of a line chart (top left in the figures), a detail view (bottom left in the figures) and an overview (bottom right in the figures) consisting of juxtaposed timelines, and a control panel (top right in the figures). The line chart and the detail view share a horizontal time axis.  

\begin{figure*}[!ht]
 \centering
 \includegraphics[width=\textwidth]{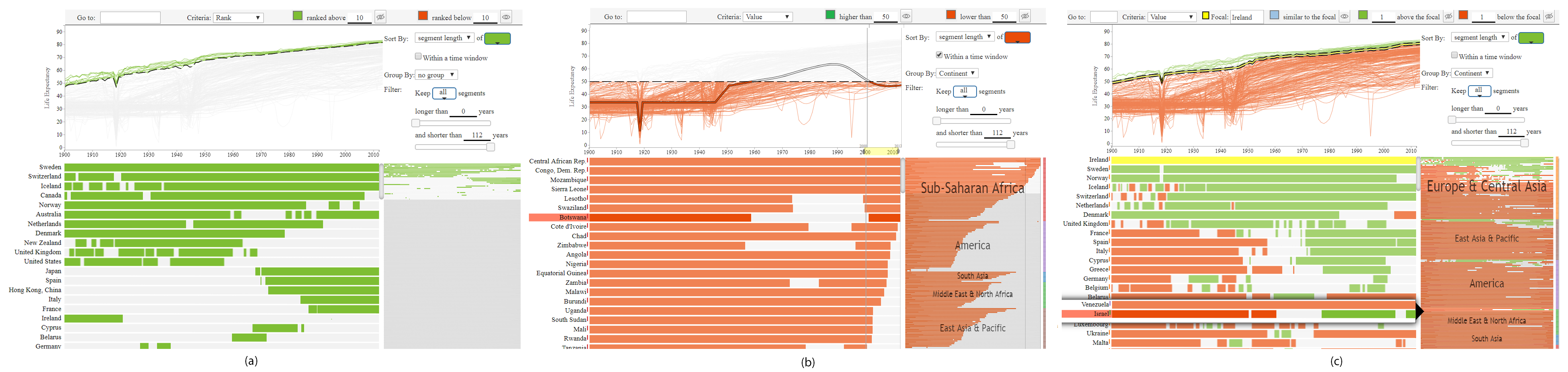}
 \vspace{-2.5em}
\caption{(a) The answers to Q1 are highlighted in green. (b) The answers to Q2 are highlighted in red. The countries are sorted by continent and then by the total length of red bars between 2000 and 2012. The time window is highlighted in yellow on the time axis. (c) The answers to Q3 are highlighted in green/red. A preview of Israel is overlaid on the detail view.}
 \label{fig_compare}
 \vspace{-1.5em}
\end{figure*}

\subsection{Line Chart and Dynamic Queries}
The line chart displays the whole dataset in the original values intuitively. The query results of interest are highlighted in colors and the rest of the data is kept gray for context.  

A set of range queries are supported in TimePool on not only original values but also derived values, including {\bf ranking} (the relative position of a data case in the original value space at a moment), {\bf percentage change} (the relative gain/loss of the value at a moment compared to the value at a given time period ago), {\bf net change} (the gain/loss of the value at a moment compared to the value at a given time period ago), and {\bf variance} (the average of the squared differences from the mean over a time window centered at a moment). The thresholds used in the queries can be constants or variables. ``Ranking $\leq$ 10'' (Fig. \ref{fig_compare}(a)) and ``life expectancy $\leq$ 50 years'' (Fig. \ref{fig_compare}(b)) are examples with constant thresholds. ``Life expectancy $\geq$ average life expectancy + 10’’(Fig. \ref{fig_le}(a)) and ``life expectancy $\leq$ life expectancy of Ireland’’ (Fig. \ref{fig_compare}(c)) are examples with variable thresholds since the average life expectancy and the life expectancy of Ireland change every year. 

TimePool allows users to {\bf directly manipulate thresholds} on the line chart whenever it is possible and provides instant visual feedback. In particular, thresholds for queries such as ``life expectancy $\leq$ 50 years'' can be represented as horizontal threshold lines in the line chart (see Fig. \ref{fig_compare}(b)). Thresholds for queries such as ``Ranking $\leq$ 10'' and ``Life expectancy $\geq$ average life expectancy + 10’’ can be represented by a threshold curve connecting the threshold values along the time axis (see Fig. \ref{fig_compare}(a) and Fig. \ref{fig_le}(a), respectively). Users can directly drag and drop the threshold lines/curves to get instant updates on the query results from the line chart and the juxtaposed timeline views. For queries related to variance and changes, users can interactively change the thresholds via sliders and get instant visual feedback from the views. 

TimePool supports {\bf two-range queries} by using one threshold (Fig. \ref{fig_le}(b) and Fig. \ref{fig_compare}(a)-(c)) and {\bf three-range queries} by using two thresholds (Fig. \ref{fig_le}(a)). Users can {\bf assign colors} or {\bf show/hide} results in one or more ranges of interest using color buttons or eye buttons tied to the ranges (i.e. green for ``Life expectancy $\geq$ average life expectancy + 10’’ and red for ``Life expectancy $\leq$ average life expectancy - 10’’ in Fig. \ref{fig_le}(a) ).

\subsection{Juxtaposed Timeline Views}

TimePool provides two juxtaposed timeline views to present the query results. The {\bf detail view} consists of juxtaposed timelines with a constant height. Each timeline represents a data case. It is colored at a time step if the data case falls into a result range to which users have assigned a color. Otherwise, it is displayed in gray. The {\bf overview} is a zoomed-out version of the detail view, displaying the timelines at a reduced height without text labels. The two views are coordinated. The data cases currently visible in the detail view are highlighted in the overview. Users can drag a scrolling bar or double-click the overview so as to display the data cases with interesting patterns observed in the overview. 

TimePool provides a {\bf sorting} function to organize data cases based on query results in the timeline views. It helps reveal patterns and allows users to access the most interesting data cases in the detail view without scrolling the screen. In particular, users can select a color from a {\bf sort by} widget to sort the data cases by the total length of timeline segments with that color. Besides sorting data cases in the whole time range of a dataset (Fig. \ref{fig_compare}(a)), users can sort them based on query results in a selected {\bf time window} (Fig. \ref{fig_compare}(b)), which can be interactively set up by users. This allows users to find insights related to a time period of interest. 

Data cases may be organized into multiple categories. Inspired by Timeline Tree \cite{burch2008timeline}, TimePool provides a {\bf group mode} to reveal patterns related to the categories. In this mode, the timelines are sorted first by category and then by the sorting criteria. The categories are also sorted by their average values of the sorting criteria. In Fig. \ref{fig_le}(a)(b) and Fig. \ref{fig_compare}(b)(c), countries are grouped by continent. A color legend is displayed on the right of each timeline in the overview to distinguish different groups.

Queries may generate very long/short colored timeline segments (i.e. after a data case falls into a result range of interest for a very long/short period of time, it falls out of it). {\bf Filtering} is provided to remove long or short segments so that users can focus on transient or long-lasting patterns. Users can {\bf select} a data case by clicking it on any views. With the shift key pressed, users can select all those cases between two clicked data cases they click in the detail view. It is called a {\bf bulk selection}. The selected data cases are highlighted in all the views. Users can also {\bf hover} the mouse over a data case to highlight it in all the views. If a data case under the mouse is not currently visible in the detail view, a timeline preview of that data case and its neighbors will be overlaid on the detail view (see Fig. \ref{fig_compare}(c)).  

\section{Example Scenario}

In this scenario, Jeff, a public health analyst, used TimePool to investigate the WLE dataset, seeking answers to Q1, Q2, and Q3. To address Q1, Jeff set the criterion as ``Rank'' with a threshold of 10. To focus on the top 10, Jeff colored line charts and timelines (lines in short) of the top 10 countries in green and hid countries without green lines (Fig. \ref{fig_compare}(a)). By default, the countries were sorted by the total length of their green lines in the timeline views, revealing an overall ranking of the countries. From the top of the detail view, Jeff observed that Sweden was always among the top 10, and Switzerland was in most of the time, while Canada was no longer among the top 10 after 2006. At the lower part of the detail view, he noticed that the top 10 countries dramatically changed after the 1960s when Japan, Spain, and Hong Kong, China replaced New Zealand, the United Kingdom, and the United States. It triggered his interest in continent-related patterns. He thus grouped the results by continent and found that, although there were much more top 10 countries in Europe \& Central Asia than in East Asia \& Pacific in history, the difference had been gradually reduced after the 1960s. As of 2012, there were four countries in East Asia \& Pacific among the top 10, versus six countries in Europe \& Central Asia.    

To answer Q2, Jeff set the criterion as ``Value'' with a threshold of 50 years. To highlight countries of interest and reveal group patterns, he colored lines when life expectancies were shorter than 50 years in red, hid countries without red lines, and grouped the countries by continent. Since he was more concerned about countries below the threshold in the 21 century, he sorted the countries by the total length of their red lines between 2000 and 2012. From Fig. \ref{fig_compare}(b), he gained valuable insights. In the overview, he observed that all continents, except Sub-Saharan Africa, surpassed the threshold between 2000 and 2012, indicating an overall improvement in life expectancy. He then conducted further analysis of Sub-Saharan Africa using the detail view. From the top of the view, he observed that several Sub-Saharan African countries were consistently below the threshold. Just below these countries, he observed a distinct pattern where several nations initially achieved the threshold around the 1960s or 1970s but subsequently experienced a decline, falling below the threshold again in the 1990s. This pattern highlighted the challenges faced by these countries in sustaining and improving life expectancy over time.

To tackle Q3, Jeff set Ireland as an ego and used the Ego Mode to compare it with the rest of the world. He colored lines in green/red when a country had a life expectancy one year higher/lower than Ireland and grouped the countries by continent (see Fig. \ref{fig_compare}(c). From the detail view, Jeff observed that Sweden was the only country that consistently had a higher life expectancy than Ireland and that multiple countries in Europe \& Central Asia surpassed Ireland after the 1960s. From the overview, he observed similar patterns in East Asia \& Pacific and America. He also noticed a solitary country with green lines in Middle East \& North Africa. Intrigued, he hovered the mouse over the segment, revealing that the country was Israel. This observation highlighted Israel's position in the region as a country with higher life expectancies than its neighbors.

Many other insights were discovered from the WLF dataset. Fig. \ref{fig_le} shows several examples. We also conducted a case study with a Beijing citizen to explore a dataset recording PM 2.5 air pollutant levels in Beijing. Many insights were discovered in the study, which can be seen in the supplementary videos.  

\section{User Study}
To evaluate how the timelines + line chart design of TimePool improves the effectiveness and efficiency in the target tasks, we conducted a user study where TimePool was compared with a baseline system (Baseline, in short). In Baseline, a line chart occupied the total screen space of TimePool. All query functions available in TimePool were provided. To enable grouping, a drop-down list was provided so that users could select one or more groups to highlight them by hiding other data cases. A search box and hovering function were provided so that users could select and examine individual data cases for detail. Two task sets about the WLE dataset
were used: T1 (Extreme): ``Who is the newest member of the top 10 (set 1)/15(set 2)?''; T2 (Condition): ``Identify at least 3 countries that were not affected by World War II (1939-1945) but suffered at some time after it (set 1)/Identify at least 3 countries that suffered three decreasing periods (set 2)''; and T3 (Comparison): ``Which country exceeded Korea (set 1)/the United States (set 2) most recently?'' 

Twelve seniors and graduate CS students taking the same visual analytics class conducted the study one by one. Each student used both TimePool and Baseline systems in sequential order and with different task sets. The order and the task sets were balanced between the systems. For each system, the subjects conducted the tasks independently after a training session, which included a live demo and 15-minute hands-on practice on a test task set (similar to the task set) on a dataset with the GDP values of 197 countries from 1990 to 2012 \cite{rosling_gapminder_2006}. The completion time and accuracy for the tasks are reported in Table 1. The subjects consistently preferred TimePool in their comments, such as ``TimePool is less time-consuming and more direct to get the answer."
``The selection (in TimePool) is far easier. You would have everything in one glance."
``Having the bar as below is easier to see where the green is and orange is for that specific country. In Baseline is very hard to do that because you have to go through each of the lines."
``TimePool is more intuitive and it is easier to understand and check the data, because this system represents very neatly.''

\vspace{-.5em}
\begin{table}[h]
\centering
\begin{tabular}{|l|l|l|}
\hline
 &Time (sec)&Accuracy\\
\hline
T1: Ti/Bs&43 (26)/124 (30)&92\% (0.29)/50\% (0.52)\\
\hline
T2: Ti/Bs&88 (42)/185 (69)&91\% (0.22)/74\% (0.35)\\
\hline
T3: Ti/Bs&73 (27)/178 (98)&96\% (0.14)/63\% (0.48)\\
\hline
\end{tabular}
\label{tab2}
\vspace{0.5em}
\caption{Averages and STDEV of results. Ti: TimePool; Bs: Baseline}
\vspace{-1.5em}
\end{table}

\section{Feedback from Domain Experts}
We interviewed 18 domain experts to learn their opinions on the usefulness and usability of TimePool. The domain experts include 15 research and teaching professors at all ranks, a postdoc, and two Mechanical Engineers. Their expertise fields include Health Informatics, Educational Research/Evaluation, Information Retrieval, HCI, Transportation Engineering, Parallel and high-performance computing, Natural Language Processing, and Machine Learning. They all work with time series data. 



One of the authors conducted the interviews via Zoom meetings where the domain experts were interviewed one by one or in groups of two. Each meeting lasted 20-60 minutes. All meetings started with a 10-minute live demo of TimePool using the WLE dataset and followed by free discussions on thoughts and suggestions. A Google Form questionnaire was distributed to the experts to return their comments and rate right after the demonstration. After the meetings, the domain experts were asked to rate the system, explain their ratings, and comment on potential usages of TimePool in
their fields of expertise through a Google form.

These domain experts rated TimePool an average of 8.7, 9.0, and 8.9 out of 10 for its usefulness, intuitiveness, and ease of learning. They made many positive comments such as: ``Very useful to get initial insights from the data.’’ ``The tool can help us filter out useless data and only show the information that is interesting, and thus significantly helping users navigate the performance data and pinpoint the exact places that have performance problems.’’ ``It provides an intuitive and efficient way to learn the pattern hidden in a large number of complex time series datasets.’’ ``I can quickly understand what each sub-graph means without too many explanations.’’ ``The design is self-explanatory.’’ ``Lots of the interactions I saw were very intuitive and natural.’’ ``It only took a few minutes of looking at the tool to understand its capabilities.’’

One domain expert had a slight concern about the learning curve:  ``I feel at the beginning, users may have some learning burden or learning curve. After a while, it should be fine with users.’’ Multiple domain experts commented on the limitations of TimePool in usability: ``It is useful. But more features can be added to make it available to a bigger audience.’’ The wish list of features discussed in the comments include: data input from varying data sources, such as NoSQL databases, preprocessing, annotation, reporting, views for visualizing selected data in full detail, customized queries associated with signal processing, and automated grouping using machine learning or other techniques. In addition, multiple domain experts hoped TimePool could be extended to support visualization of multidimensional time series data.  

The domain experts suggested many potential applications of TimePool in their fields of expertise, such as analyzing data from COVID-19, student performance, high-performance computing, and traffic studies. For example, a researcher on Health Informatics commented that ``The ability to drill the data (query) will provide interesting visualizations and insight into the data (about COVID-19) when grouped by city, country, continent over the time period.’’ A researcher on Transportation Engineering commented that ``The tool can be used to visually display the time of day, day-to-day, and seasonal changes as well as the yearly changes and comparisons. This can be
very helpful to facilitate the transportation infrastructure management
and roadway improvement project decisions.’’

\section{Conclusion and Future Work}
 
This paper introduces TimePool, a new visualization prototype for addressing ``which and when'' tasks in univariate time series analysis. At its core is the idea of conducting queries at individual time steps and
comprehensively inspecting the ``which and when'' results in a familiar line graph and complementary timelines. TimePool allows users to effectively and efficiently conduct low-level analytical tasks about extremes, conditions, and comparisons. A rich set of interactions are provided for exploratory analysis. Users can interactively change query types, thresholds, and focus time span, thereby inspiring new insights, tasks, and hypotheses. Additionally, TimePool facilitates comparative analyses by enabling data grouping. A demo version of this prototype written in JavaScript can be accessed at \url{https://cs.appstate.edu/fengt/TimePool/}.

The example scenario and feedback from domain experts revealed the potential of TimePool in conducting exploratory analysis of time series data. The user study showed that the line chart + timelines design of TimePool achieved better performance than traditional line charts in speed and accurary when conducting ``which and when'' tasks. As commented by the users, TimePool provides an intuitive and efficient way to analyze time series data and can be applied to a wide range of applications. We need to point out that TimePool is focused on ``which and when'' tasks for univariate time series, and thus its applicability is limited. Its scalability for large datasets also needs to be improved to ensure smooth user experiences. 

\bibliographystyle{abbrv-doi}

\bibliography{template}
\end{document}